\documentclass{rspublic}

\begin{document}

\title[Mechanical heating by active galaxies]{Mechanical heating by active galaxies}

\author[M. C. Begelman \& M. Ruszkowski]{Mitchell C. Begelman and Mateusz Ruszkowski}

\affiliation{JILA, 440 UCB, University of Colorado, Boulder, CO 80309-0440, USA}

\label{firstpage}

\maketitle

\begin{abstract}{intracluster medium --- active galaxies --- jets --- cooling flows --- feedback --- entropy}
Jets and winds are significant channels for energy loss from accreting black holes. These outflows mechanically heat their surroundings, through shocks as well as gentler forms of heating. We discuss recent efforts to understand the nature and distribution of mechanical heating by central AGNs in clusters of galaxies, using numerical simulations and analytic models. Specifically, we will discuss whether the relatively gentle  `effervescent heating' mechanism can compensate for radiative losses in the central regions of clusters, and account for the excess entropy observed at larger radii.
\end{abstract}

\section{Introduction}

Observational and theoretical lines of evidence suggest that accreting black holes in active galaxies release large amounts of kinetic energy to their environments. Observationally, a minority of active galaxies release the bulk of their power in the form of radio jets (Rees \textit{et al.} 1982; Begelman \textit{et al.} 1984). But more numerous radio quiet AGN may also generate much of their power in the form of fast winds, as indicated by recent spectral analyses of broad absorption line (BAL) QSOs in the optical/UV (Arav \textit{et al.} 2001; de Kool \textit{et al.} 2001) and X-ray (Reeves \textit{et al.} 2003) bands. These studies suggest that the kinetic energy in BAL outflows --- which are thought to occur in most radio quiet QSOs --- is larger than previously thought and can approach the radiation output. 

Theoretical arguments also point to the importance of the kinetic energy channel.  Numerical simulations suggest that accretion disks, which transfer angular momentum and dissipate binding energy via magnetorotational instability, may inevitably produce magnetically active coronae (Miller \& Stone 2000).  These are likely to generate outflows that are further boosted by centrifugal force (Blandford \& Payne 1982).  Unless radiation removes at least 2/3 of the liberated binding energy, very general theoretical arguments indicate that rotating accretion flows {\it must} lose mass.  The physical reason is that viscous stresses transport energy outward, in addition to angular momentum.  If radiation does not remove most of this energy, then a substantial portion of the gas in the flow will gain enough energy to become unbound (Narayan \& Yi 1995; Blandford \& Begelman 1999, 2004). Blandford (this volume) discusses this effect and its consequences in more detail. While it may sometimes be possible to tune the system so that the gas circulates without escaping, any excess dissipation (i.e., increase of entropy) near a free surface of the flow will lead to outflow.  There are several possible sources of such dissipation, including magnetic reconnection, shocks, radiative transport, and the magnetocentrifugal couple mentioned above. If radiative losses are very inefficient, outflows can remove all but a small fraction of the matter supplied at large radii.   

While it is probably safe to say that some substantial release of kinetic energy always accompanies accretion, we are not yet able to predict its magnitude, nor how it compares with radiative losses. During the growth of a supermassive black hole to mass $M_{BH}$, the integrated kinetic energy output could be comparable to the total release of binding energy, $\sim 0.1 M_{BH} c^2$.  It is unlikely to be much smaller than a tenth that amount. Even with a kinetic energy output of $\sim 0.01 M_{BH} c^2$, the effects on the environment can be dramatic.  At a kinetic energy conversion efficiency of $\varepsilon_{KE} c^2$ per unit of accreted mass, an accreting black hole liberates $10^{19} (\varepsilon_{KE}/0.01)$ ergs per gram.  In a galactic bulge with a velocity dispersion of $200 \sigma_{200}$ km s$^{-1}$, the accretion of one gram liberates enough energy to accelerate $2 \times 10^4 (\varepsilon_{KE}/0.01) \sigma_{200}^2 $ grams to escape speed --- provided that most of the energy goes into acceleration.  Since the typical ratio of black-hole mass to galactic bulge mass is $> 10^{-3}$ (H\"aring \& Rix 2004), feedback from a supermassive black hole growing toward its final mass could easily exceed the binding energy of its host galaxy's bulge.  Such feedback has been invoked in various models to explain the $M_{BH}-M_{bulge}$ and $M_{BH}-\sigma$ correlations (e.g., Silk \& Rees 1998; Blandford 1999; Fabian 1999).

The effects of injecting a certain amount of kinetic energy into the environment of a supermassive black hole depend not only on the amount of energy injected, but also on the temporal and spatial characteristics of the injection process and the structure of the ambient medium. Explosive injection of a large amount of energy in a short time --- which might lead to intense, centrally concentrated heating at a shock --- will have very different consequences from gradual, intermittent, or spatially distributed injection, which would allow the surrounding medium to adjust.  Whether mechanical heating is partially offset by radiative cooling is an important factor, as is the presence of absence of small-scale density inhomogeneities.  The speed of a shock or sound wave propagating through a medium with a  `cloudy' thermal phase structure will be highest in the phase with the lowest density (the intercloud medium).  Dense regions will be overrun and left behind by the front, as first pointed out by McKee \& Ostriker (1977) in connection with supernova blast waves propagating into the interstellar medium. Consequently, most of the energy goes into the gas which has the lowest density (and is the hottest) to begin with.  The global geometric structure of the ambient gas is important as well.   Since a wind or hot bubble emanating from an AGN will tend to follow the path of least resistance, a disk-like structure can lead to a  `blowout' of hot gas along the axis, leaving the disk intact.  Anisotropic injection of the kinetic energy, e.g., in a pair of jets or equatorial, can likewise affect certain regions while sparing others. 

Given these complications, it is especially desirable to find a  `laboratory' where one can study the details of mechanical energy injections by active galaxies in a specific context.  Clusters of galaxies can serve this role well.  In the remainder of this paper we will discuss how recent X-ray observations of the intracluster medium (ICM) suggest a particular mode of heating by AGNs, which appears to be susceptible to theoretical modeling.  While mechanical heating is likely to be the only important form of AGN feedback into the hot, highly ionized atmospheres of galaxy clusters, we stress that other forms of energy injection, particularly radiative heating, must occur as well.  These effects are can be particularly important on smaller scales (i.e., the interstellar medium of the host galaxy) and in less highly ionized environments; they are discussed by Ostriker (this volume).

\section{The nature of heating in clusters}

Two major pieces of evidence point to mechanical heating by AGN as an important energy source in the ICM.  First, there is the absence of  `cooling flows'. Radiative cooling timescales in the central regions of clusters are often much shorter than the Hubble time. Initially, this led to suggestions that the intracluster medium (ICM) is flowing into the cluster center at rates of up to 1000 $M_{\odot}/$yr.  However, {\it XMM Newton} and {\it Chandra} observations suggest that the actual inflow rates are much smaller than expected, indicating that some mechanism compensates for the cooling (e.g., Fabian \textit{et al.} 2000, 2002; McNamara \textit{et al.} 2000; Blanton \textit{et al.} 2001; Churazov \textit{et al.} 2002). The possibility that active galaxies provide the heating is supported by observations that $\sim 70\%$ of cD galaxies show evidence for active radio sources (Burns 1990), and that the  ensemble-averaged power from radio galaxies may be adequate to offset the mean level of cooling (Peres \textit{et al.} 1998; B\"ohringer \textit{et al.} 2002).  An advantage of the AGN heating model over other models (e.g., thermal conduction from large radii) is that the heating is supplied near the cluster center where the cooling flow problem is most acute. AGN heating may explain why the gas temperature, while declining towards cluster centers, does not drop by more than a factor $\sim 2-3$ between the cooling radius and the cluster center (Allen \textit{et al.} 2001; Fabian \textit{et al.} 2001; Peterson \textit{et al.} 2001, 2003).  

The second piece of evidence is the excess entropy found in clusters. Cluster X-ray luminosities and gas masses increase with temperature more steeply than predicted by hierarchical merging models (Markevitch 1998; Nevalainen \textit{et al.} 2000).  In other words, the atmospheres in less massive clusters and groups are hotter than they should be, given the gravitational interactions that assembled them. These correlations apply to regions of clusters well outside the cooling radius, as well as to clusters without cooling cores.
They can be interpreted as evidence for an entropy  `floor' (Lloyd-Davies et al.~2000) or a systematic excess of entropy (Ponman \textit{et al.} 2003); the most plausible explanation appears to be AGN heating before or during cluster assembly (e.g., Valageas \& Silk 1999; Nath \& Roychowdhury 2002; McCarthy et al. 2002, and references therein).  

Both pieces of evidence suggest that the heating, whatever its cause, must be both widely distributed and gentle. Radiatively cooling gas in clusters is prone to thermal instability, since the cooling rate increases rapidly with density.  To avoid a  `cooling catastrophe', heat must be spread evenly though the ICM, and especially targeted at regions with large density gradients.  Measurements of excess entropy also show that the heat must be spread over a range of radii, extending out to half the cluster virial radius.  That the central heat source is relatively gentle in cooling flow clusters is suggested by the absence of X-ray emitting shocks bounding radio lobes (Fabian \textit{et al.} 2000) and the fact that cluster cores appear to have positive radial entropy gradients, i.e., they are convectively stable (David \textit{et al.} 2001; B\"ohringer \textit{et al.} 2002).

Can a centrally located AGN provide mechanical heating that is gentle, yet spreads widely through the cluster? At first glance it seems unlikely.  Powerful radio galaxies, like Cygnus A, produce overpressured cocoons that expand supersonically into their surroundings.  After a transient phase dominated by the momentum flux in the jets, cocoons resemble spherical, supersonic stellar wind bubbles (Begelman \& Cioffi 1989). The evolution of the bubble can be described approximately by a self-similar model in which the internal and kinetic energy are comparable, and share the integrated energy output of the wind.  The speed of expansion is 
\begin{equation}
v\sim \left({L_j\over \rho}\right)^{1/5} R^{-2/3},
\end{equation}
where $L_j$ is the power of the jets, $\rho$ is the ambient density, and $R$ is the radius of the shock.  The supersonic expansion phase ends when the expansion speed drops below the sound speed in the ambient medium. This occurs at a radius
\begin{equation}
R_{\rm sonic}\sim 5 \left({\langle L_{43} \rangle \over n}\right)^{1/2} T_{\rm keV}^{-1/4} \ {\rm kpc} , 
\end{equation}
where $\langle L_{43} \rangle$ is the time-averaged jet power in units of $10^{43}$ erg s$^{-1}$, $n$ is the ambient particle density in units of cm$^{-3}$, and $T_{\rm keV}$ is the ambient temperature in units of keV.  Thereafter the evolution is dominated by buoyancy (Gull \& Northover 1973).  We have chosen fiducial parameters that are fairly typical of conditions at the centers of rich clusters --- note how small $R_{\rm sonic}$ is, compared to a typical cluster core radius, or even the core radius of the host galaxy.  Cygnus A, which has been expanding for several million years, is hundreds of kpc across, and is still overpressured by a factor $\sim 2-3$ with respect to the ambient medium, is the exception rather than the rule.  It is a very powerful source expanding into a relatively tenuous ambient medium (Smith \textit{et al.} 2002). During most of the evolution of clusters we can expect the energy injection to be in the buoyant regime, and the heating therefore relatively gentle.

A clue to the widespread distribution of the heat comes from the apparent  immiscibility of the hot (possibly relativistic) plasma injected by the jets and the thermal gas of the ICM. It has been known since the time of {\it ROSAT} (B\"ohringer \textit{et al.} 1993) that the plasma in radio lobes can displace cooler thermal gas, creating  holes in the X-ray emission.  More sensitive {\it Chandra} images have shown not only how common such holes are, but also how long they can persist. In particular, numerous examples of  `ghost cavities' have been found (e.g.,McNamara \textit{et al.} 2001; Johnstone \textit{et al.} 2002; Mazzotta \textit{et al.} 2002), where the X-ray deficit persists but the compensating radio emission is either absent or too weak to detect. These are presumably buoyant bubbles left over from earlier epochs of activity.  

The persistence of highly buoyant bubbles implies that energy can be transported to large radii, despite the convective stability of the ICM. The Schwarzschild criterion refers to heat transport by marginally buoyant fluid elements, not the highly buoyant bubbles that appear to be present. We therefore obtain the following description of how the ICM can be heated by a central AGN:

\bi
\item Pockets of very buoyant gas rise subsonically through the ICM pressure gradient. A large density contrast is maintained between the buoyant gas and its surroundings, i.e., there is little mixing.

\item The buoyant gas does $pdV$ work on the ICM as it rises and expands.  This work goes initially into a combination of kinetic energy, internal energy, and gravitational potential energy (e.g., sound waves, g-modes, and internal waves).

\item The energy transferred to the ICM energy is converted to heat by damping and/or mixing.  
\ei

We call this process \textit{effervescent heating} (Begelman 2001; Ruszkowski \& Begelman 2002). 

\section{Modeling mechanical heating}

\subsection{Numerical simulations}

The most direct way to study such a complex set of fluid dynamical processes is through numerical simulations.  A variety of 2D and 3D simulations have now been published, addressing the heating of the ICM by buoyant plumes. These calculations illustrate several of the requirements for gentle, distributed heating to occur:  

\bi
\item The plumes spread laterally as well as radially, a result of the  `mushroom cloud' effect (Churazov \textit{et al.} 2001; Ruszkowski \textit{et al.} 2004). This is necessary in order to distribute the energy in solid angle, if it is initially injected by jets.

\item The bubbles persist long after the observable radio lobes have faded (Reynolds \textit{et al.} 2002; Basson \& Alexander 2002; Br\"uggen \textit{et al.} 2002), and may drift far from the initial injection axis.  Since radio emission fades rapidly once the bubbles begin to rise, this provides a straightforward explanation for  `ghost cavities' (M. Ruszkowski, C. Kaiser \& M. Begelman 2003, unpublished work).

\item The plumes entrain and lift material from the cluster core, giving rise to the observed cool rims around the radio lobes (Brighenti \& Mathews 2002; Br\"uggen 2003).

\item The rising bubbles and associated mixing do not necessarily smear out abundance gradients in the ICM (Br\"uggen 2002).

\item Rising and expanding bubbles increase the potential, thermal and kinetic energy of the ICM (Reynolds \textit{et al.} 2002; Quilis \textit{et al.} 2001; Churazov \textit{et al.} 2002; Br\"uggen \& Kaiser 2002).  If the AGN is intermittent they can generate sound waves (Ruszkowski \textit{et al.} 2004), which may have been detected in the Perseus (Fabian \textit{et al.} 2003) and the Virgo (Forman \textit{et al.} 2003) clusters.
\ei

\subsection{Analytic modeling}

Analytic `toy' models are another useful way to model the principal features of mechanical heating by AGNs.  We consider the $pdV$ work done by a spherically symmetric ensemble of bubbles rising subsonically through the pressure gradient. Since the timescale for the bubbles to cross the cluster (of order the free-fall time) is much shorter than the cooling timescale, the flux of bubble energy through the ICM approaches a steady state, implying that details of the energy injection process --- such as the number flux of bubbles (e.g., one big one or many small ones), the bubble size, filling factor, and rate of rise --- do not affect the mean heating rate. If we assume that the $pdV$ work is dissipated within a pressure scale height of where it is generated, we can devise an average volume heating rate for the ICM, as a function of radius (Begelman 2001):  
\begin{equation}
\label{effer}
{\cal H} \sim {\langle L_b \rangle \over 4\pi r^3}\left({p\over p_0}\right)^{1/4}\left|{d\ln p \over d\ln r}\right| .
\end{equation}
In eq.~(\ref{effer}), $\langle L_b \rangle$ is the time-averaged power output of the AGN going into bubbles, $p(r)$ is the pressure inside the bubbles (and $p_0$ is the pressure where the bubbles are formed), and the exponent $1/4$ equals $(\gamma - 1)/\gamma$ for a relativistic plasma (the exponent would be 2/5 for a nonrelativistic gas).  A major assumption of the model, that the $pdV$ work is absorbed and converted to heat within a pressure scale height, will have to be assessed using numerical simulations and studies of the microphysics of cluster gas. We have also assumed that the energy is spread evenly over $4\pi$ sr, a likely consequence of buoyancy . 

The most important property of the above effervescent heating rate is its proportionality to the pressure gradient (among other factors), since this determines the rate at which $pdV$ work is done as the bubbles rise.  Since thermal gas that suffers excess cooling will develop a slightly higher pressure gradient, the effervescent heating mechanism targets exactly those regions where cooling is strongest.  Therefore, it has the potential to stabilize radiative cooling (Begelman 2001). This potential is borne out in 1D, time-dependent numerical simulations of a cooling core (Ruszkowski \& Begelman 2002), which show that the flow settles down to a steady state that resembles observed clusters, for reasonable parameters and without fine-tuning the initial or boundary conditions. Although these models include conduction (at 23\% of the Spitzer rate), which may be necessary for global stability (Kim \& Narayan 2003), the heating is overwhelmingly dominated  by the AGN at all radii. The mass inflow rate through the inner boundary, which determines the AGN feedback in these simulations, stabilizes to a reasonable value far below that predicted by cooling flow models. 

According to the analytic models, a large fraction of the injected energy reaches radii much larger than the cooling radius.  Roychowdhury \textit{et al.} (2004) have used these models to study whether AGN heating can explain the excess entropy of cluster gas.  We find that the distributed heating rate given by eq. (3.1) can reproduce simultaneously the luminosity--temperature correlations measured at two different radii (roughly 0.5 and 0.08 of the virial radius), for cluster temperatures between 1 and 10 keV. These results appear to require a high efficiency of kinetic energy production ($\varepsilon_{KE}$) during the growth of the black hole as well as a black hole mass (or sum of black hole masses) roughly proportional to the virial mass of the cluster.  If we assume $M_{BH} \approx 0.0015 M_{bulge}$ for each galactic bulge in the cluster (H\"aring \& Rix 2004), and further assume that bulges comprise a fraction $0.01 f_{-2}$ of the cluster's virial mass, then the entropy measurements require $\varepsilon_{KE} f_{-2}\approx 0.05$.  Realistically this represents an upper limit, since it was calculated assuming that all of the heating occurs at low redshift, i.e., after each cluster has fully formed.  Since most black hole growth appears to have occurred by accretion during the `quasar era' at $z \sim 2-4$ (Yu \& Tremaine 2002), the entropy we see today was probably injected into the lower-mass precursors of modern clusters, which had lower virial temperatures.  According to the second law of thermodynamics, the heat input required to produce a given change in entropy is proportional to temperature; therefore, the required value of $\varepsilon_{KE} f_{-2}$ is probably smaller than the one stated above.

\section{Conclusions}

There is growing evidence, from observations as well as theory, that active galaxies can provide widespread mechanical heating of their environments.  In clusters, this heating apparently occurs in a rather gentle fashion, and is driven by buoyancy.  Simulations of buoyant plumes show significant and fast lateral spreading, generation of sound waves, cool rims of entrained gas surrounding the hot bubbles and a mismatch between X-ray and radio emission, resulting in `ghost cavities'.  The evenness and spatial distribution of heating may be adequate to balance cooling globally, prevent cooling catastrophes, and thus quench so-called `cooling flows'.  Energy spreading to larger radii during cluster assembly might be able to account for the observed entropy excesses in present-day clusters.  

While the initial results are promising, numerical simulations have a long way to go before they can adequately the represent the physics of AGN heating.  Three-dimensional simulations are already being done, but their resolution needs to be increased in order to study the effects of mixing and thermal instability.  Magnetic fields, which have played little role in models to date, may have important effects on the dynamics, transport properties (viscosity and thermal conduction), and radio emissivity of clusters.  At the microphysical level, we need to understand why the hot (relativistic?) gas injected by active galaxies appears to mix relatively little with the ICM.  Indeed, we cannot exclude the possibility that some of the injected `cosmic rays' do stream through the thermal background.  If they couple effectively to the ICM via hydromagnetic waves, they will heat the gas in much the same fashion as expanding bubbles, as they traverse the pressure gradient (Loewenstein \textit{et al.} 1991).   

Finally, we need to better understand what sets the efficiency of kinetic energy output from black hole accretion flows, the speed and degree of collimation of the output (winds vs. narrow jets), and feedback effects that couple the evolution of the ICM to the growth rate of the black hole.  Whether (and how) such feedback fixes the $M_{BH}- M_{bulge}$ correlation by regulating the black hole mass, the galaxy mass, or both, remains to be seen.

\begin{acknowledgements}
Support for this work was provided by National Science Foundation grant AST-0307502 and the National Aeronautics and Space Administration through
{\it Chandra} Fellowship Award Number PF3-40029 issued by the {\it Chandra} X-ray Observatory Center, which is operated by the Smithsonian
Astrophysical Observatory for and on behalf of the NASA under contract NAS8-39073.  Some of the work reported here was done in collaboration with M. Br\"uggen (International University Bremen); S. Roychowdhury and B. Nath (Raman Research Institute, Bangalore); and C. Kaiser (University of Southampton).
\end{acknowledgements}

\label{lastpage}

\end{document}